**Investigating and improving student understanding of the basics of quantum computing**

Peter Hu*, Yangqiuting Li, and Chandralekha Singh
Department of Physics and Astronomy, University of Pittsburgh, Pittsburgh, PA 15260, USA

*Corresponding author, pth9@pitt.edu

**Abstract**

Quantum information science and engineering (QISE) is a rapidly developing field that leverages the skills of experts from many disciplines to utilize the potential of quantum systems in a variety of applications. It requires talent from a wide variety of traditional fields, including physics, engineering, chemistry, and computer science, to name a few. To prepare students for such opportunities, it is important to give them a strong foundation in the basics of QISE, in which quantum computing plays a central role. In this study, we discuss the development, validation, and evaluation of a QuILT, or Quantum Interactive Learning Tutorial, on the basics and applications of quantum computing. These include an overview of key quantum mechanical concepts relevant for quantum computation (including ways a quantum computer is different from a classical computer), properties of single- and multi-qubit systems, and the basics of single-qubit quantum gates. The tutorial uses guided inquiry-based teaching-learning sequences. Its development and validation involved conducting cognitive task analysis from both expert and student perspectives and using common student difficulties as a guide. For example, before engaging with the tutorial, after traditional lecture-based instruction, one reasoning primitive that was common in student responses is that a major difference between an $N$-bit classical and $N$-qubit quantum computer is that various things associated with number $N$ for a classical computer should be replaced with number $2^N$ for a quantum computer (e.g., $2^N$ qubits must be initialized and $2^N$ bits of information are obtained as the output of the computation on the quantum computer). This type of reasoning primitive also led many students to incorrectly think that there are only $N$ distinctly different states available when computation takes place on a classical computer. Research suggests that this type of reasoning primitive has its origins in students learning that quantum computers can provide exponential advantage for certain problems, e.g., Shor's algorithm for factoring products of large prime numbers and that the quantum state during the computation can be in a superposition of $2^N$ linearly independent states. The inquiry-based learning sequences in the tutorial provide scaffolding support to help students develop a functional understanding. The final version of the validated tutorial was implemented in two distinct courses offered by the physics department with slightly different student populations and broader course goals. Students' understanding was evaluated after traditional lecture-based instruction on the requisite concepts, and again after engaging with the tutorial. We analyze and discuss their improvement in performance on concepts covered in the tutorial.

**Introduction**

Quantum information science and engineering (QISE) is an exciting interdisciplinary field that has applications in quantum computing, quantum communication and networking, and quantum sensing, which are attractive to scientists and engineers for many reasons. Computer scientists and engineers are developing quantum algorithms for various problems, including ones that become impractical for classical computers to solve at large scales. For example, on a classical computer, the problem of factoring products of large prime numbers scales exponentially with the size of the prime numbers, but on a quantum computer utilizing Shor's algorithm, the problem scales roughly as a polynomial instead. Given that the difficulty of factoring products of large prime numbers is at the heart of the RSA (Rivest-Shamir-Adleman) protocol used for encryption of sensitive data, the development of Shor's algorithm showing a quantum computer's exponential speed-up provided a major impetus to QISE research. As another example, Grover's algorithm also provides an advantage for quantum computers over classical computers in searching an unsorted list, this time as a quadratic speed-up. Since this type of list is often encountered in many applications, even a quadratic speed-up constitutes a major improvement. For future applications in science, physicists and chemists are also excited about the potential of quantum computers to solve important problems in their disciplines in which solving the Schrödinger equation plays an important role. The development of robust quantum bits (qubits) and scalable quantum computers demands the expertise of physicists and engineers alike. For all these reasons and more, this area of study holds a great amount of promise for students from many science and engineering disciplines interested in careers in QISE-related fields [1–6].

To ensure their success in the field, students must develop a strong foundational knowledge of qubits and quantum systems at the start of their studies, as well as a sense of the advantages that quantum computers are capable of providing over classical computers. Here we describe the development, validation and in-class implementation of a research-based Quantum Interactive Learning Tutorial (QuILT) on these topics. The tutorial was implemented in two types of courses at a large research university in the United States: one is the standard two-semester quantum mechanics (QM) course sequence for physics majors, and the other is an interdisciplinary course involving students from many fields, the only mandatory course in a new "Foundations of Quantum Computing and Quantum Information" undergraduate certificate program offered by the institution to undergraduate students across science and engineering disciplines.

Learning QM is challenging for students partly since the quantum paradigm is very different from the classical paradigm [7]. Prior research suggests that students in QM courses often share common difficulties [7–13], in a number of concepts [14–19] at all levels [20–24]. Prior work also indicates that research-validated learning tools can effectively help students develop a functional understanding [25–27], focusing on how to successfully foster student sensemaking [28,29], including through visualizations and various technology-embedded resources [30–34]. We have been researching student difficulties after traditional lecture-based instruction and using the research to guide the development of learning tools for concepts covered in undergraduate QM courses. Previous work from our group includes QuILTs on topics such as the Mach-Zehnder interferometer, the double-slit experiment, quantum key distribution, fand the Bloch sphere, among others [35–41]. Our group has also developed and validated Clicker Question Sequences on topics such as the basics of two-state systems and change of basis, quantum measurement, time-development, and measurement uncertainty of two-state quantum systems [42–45]. We find that after traditional lecture-based instruction, students may not have a strong grasp of important quantum concepts, but after further engagement, e.g., with a research-based QuILT, they may develop additional fluency with those concepts. To that end, we have developed and validated a QuILT (referred to in this paper as "the tutorial") to help students learn about the basics of quantum computing.

Given the interdisciplinary nature of the QISE field, for which this tutorial is intended, there is a need to standardize the language to be accessible and unambiguous for everyone regardless of background. Some trends have already taken hold in the field, including a distinction between interpreting phenomena "classically" as opposed to "quantumly," rather than "quantum mechanically"; speaking of "measuring qubits" as opposed to measuring a physical observable; referring to a "measurement basis"; or measuring specific states such as $|0\rangle$ or $|1\rangle$ as outcomes rather than the corresponding observables (eigenvalues). These are evolutions in linguistic construction that quantum physicists have typically not used in the past. However, as there is a one-to-one correspondence, e.g., between the eigenvalues obtained by making a measurement of an observable in a specific state, and the eigenstates that represent the standard basis as $|0\rangle$ and $|1\rangle$, it can serve as useful shorthand to say that a qubit is measured to be in the $|1\rangle$ state (or measured to be 1) to convey that a measurement made on the system yields the eigenvalue corresponding to the $|1\rangle$ state.

In this research on the development, validation, and in-class implementation of the quantum computing tutorial, we opt to use the prevailing terminology in the QISE field so that students can become familiar with the language used by their textbooks, instructors, and other professionals in their studies in this interdisciplinary field. Not coincidentally, this language also serves to express many concepts more directly and in fewer words, and thus may reduce the cognitive load to learn them, as many of the physical details are immaterial to many of the QISE contexts in which these students will apply the concepts that they learn. All that said, it is still crucial to maintain the integrity of language used to educate physicists involved in QISE who will be working to develop robust qubits and build real quantum computers, work that very much requires understanding the entire quantum physics taught in typical undergraduate and graduate physics courses [2–4,46].

**Theoretical framework**

In QM courses, whose content can be difficult for students, it is critical to consider constructivist research-based pedagogical approaches to engage students and help them learn these challenging, foundational concepts [7]. The theoretical framework that guided the development, validation and evaluation of this tutorial emphasizes that the research-based pedagogical approaches should balance efficiency and innovation, i.e., Schwartz et al.'s Preparation for Future Learning (PFL) framework [47]. In one interpretation of this framework, innovation in instructional approaches would focus on providing students opportunities to engage and struggle with novel problem-solving tasks so that they develop high order thinking skills and the ability to apply existing knowledge to novel situations (transfer of knowledge). Exploratory labs are examples of environments that are intended to maximize innovation. On the other hand, efficiency in instructional approaches can refer to efficient ways to communicate relevant knowledge to students, e.g., via lectures. The authors observe that, in most traditional classrooms, efficiency is overemphasized while innovation is typically disregarded, and suggest that balancing innovation and efficiency in learning activities can improve conceptual understanding and transfer of knowledge to new contexts [47].

The advantages of balancing innovation and efficiency are outlined by Nokes-Malach and Mestre [48], who propose mechanisms for the phenomenon of the "time for telling" in which students who have productively struggled with innovative invention tasks learn more deeply from subsequent, more efficient methods of instruction than students who did not undergo the invention tasks [49]. They hypothesized that allowing students to productively struggle through invention tasks imparts a "hidden" effectiveness in key ways. For example, the particular framing and productive struggle may prime students to operate in a more mastery-based orientation rather than a performance-based one; a mastery orientation is one in which students are interested in deeply understanding the material, while performance orientation implies students' desire to, e.g., get a good score or pass the course [50]. On the other hand, the students who only receive lecture-based instruction (an instantiation of efficiency) may use problem-solving steps via a template without comprehensively considering the reasoning for each step [48]. Thus, balancing innovation and efficiency is vital for priming all students to engage in deep sense making and learning within the time-constraints of the course.

Having outlined the benefits of balancing innovation and efficiency, the aforementioned fixation on efficiency in most solely lecture-based classrooms leads to what Schwartz et al. term "routine experts," who are fluent in a specific type of task but struggle with transferring their learning to new situations, rather than "adaptive experts" who are able to complete a wide variety of tasks requiring transfer of knowledge in their area of expertise with speed and accuracy. A pure focus on innovation leads to another issue: the "frustrated novice," who without sufficient guidance and scaffolding support, is unable to make any meaningful progress on problems in a given amount of time, ultimately experiencing few benefits of the freedom provided by innovation and discovery. Thus, Schwartz et al. conclude, developing students' expertise in a domain requires balancing of efficiency and innovation axes in the instructional design, outlining an "optimal adaptability corridor" by which they can develop competencies to become "adaptive experts" with the least amount of wasted effort [47].

Inspired by this framework, the tutorial on the basics of quantum computing strives to balance innovation and efficiency using guided inquiry-based teaching-learning sequences that build on one another as well as on student prior knowledge to help them develop a functional understanding. The development and validation of our tutorial involves conducting cognitive task analysis from both expert and student perspectives. From the expert perspective, we delineate a fine-grained scope of content and its relevance to our learning objectives through discussions and the iteration of ideas among ourselves many times, as well as consulting multiple faculty members experienced in teaching QM and related fields (e.g., solid state physics). The cognitive task analysis from the student perspective involves interviewing students so that student difficulties can be used as a guide and the teaching-learning sequences in the tutorial provide appropriate scaffolding at a level that balances efficiency and innovation. The cognitive task analysis from the student perspective is valuable to minimize expert blind spots.

**Methods**

*Development and validation*

To use common student difficulties with underlying concepts as a guide and balance innovation and efficiency in the tutorial, student difficulties were investigated over many years, both formally and informally, including via responses to open-ended questions asked on exams in an undergraduate QM course following traditional lecture-based instruction on the requisite concepts. These difficulties, ascertained through discussion and interactions with undergraduate students who had been instructed on the appropriate QM and QISE content, were then used as a guide over the subsequent development and validation of the tutorial.

After completion of an early draft of the tutorial, pre-test and post-test, four graduate students were invited to work through all three over the course of several consecutive think-aloud interviews completed within one week, spanning a total of roughly fifteen hours. With only a limited pool of students possessing the required level of background knowledge, soliciting undergraduate students during the time frame of the validation and implementation of the tutorial (and the corresponding pre- and post-test) would run the risk of some students working through the tutorial content both inside and outside of class. Even though the tutorial is intended for the undergraduate courses in which QISE concepts are covered, the graduate students provided an appropriate group for these interviews with their more cohesive foundational QM knowledge (having taken graduate-level QM). In particular, although these graduate students did not have explicit prior course on the QISE basics, their background enabled them to effectively engage with the material, especially in conjunction with the tutorial's extensive scaffolding. Their feedback was used to gauge overall flow and whether the tutorial was at the appropriate level, as well as to identify blind spots, and their suggestions were incorporated into the subsequent versions of the tutorial.

Throughout, the tutorial was repeatedly iterated with the graduate student interviewees' feedback, through constant discussions among the authors, and with the input of four additional faculty members experienced in teaching QM and solid-state physics. The tutorial is included in Ref. [51] and the Supplemental Material for convenience. Developed and validated alongside the tutorial at the same time, using the same process and involving the same people, were a pre-test and post-test to evaluate students' understanding of the underlying concepts. The two tests contained isomorphic questions with minor changes to some details such as given states. The post-test versions of these questions are provided in Appendix A.

*Structure of the tutorial*

The learning objectives for the tutorial are listed in Table 1.

**Table 1.** Learning objectives for the tutorial.

| Learning objective (students should be able to do the following): | Corresponding pre-test/post-test question |
|---|---|
| Identify possible states for 2-bit/qubit systems | 1a, 1b |
| Identify examples of what can be used as a bit or qubit | 1c |
| Identify possible outcomes of measurement as well as calculate probabilities of measuring each outcome in a given state | 1d, 2a, 3a, 3b, 3c, 7a, 7b 7c, 9a, 9b, 9c |
| Identify whether there are $N$ or $2^N$ of a given quantity involved in classical vs. quantum computers (e.g., describe that measurement collapses the state so that only $N$ bits of information are obtained as output of computation even for an $N$-qubit quantum computer; only $N$ | 4a, 4b, 4c, 5, 6 |
| qubits must be initialized in a quantum computer; a quantum state can in general evolve as a superposition of $2^N$ states but a classical computer which also has $2^N$ distinct states can only be in one of those states at a given time, etc.) | |
| Define and identify examples of superposition states and entangled states and be able to contrast the phenomena of superposition and entanglement | 4d, 8a, 8b, 8c, 8d |
| Describe the role and action of a single-qubit quantum gate (describe how these gates can be used to change a state) | 2b, 3c, 3d |
| Specify the bra state corresponding to a given ket state (e.g., $\lvert\chi\rangle = a\lvert0\rangle + b\lvert1\rangle$ where $a$ and $b$ are in general complex) | 10a |
| Calculate an outer product and be able to apply an outer product (operator) on a given state to find the outcome | 10b, 10c |
| Write a single-qubit gate in Dirac notation | 11 |

The tutorial consists of multiple-choice questions, hypothetical student discussions, mathematical exercises, plentiful invitations for students to explain their reasoning, and checkpoints to keep them oriented on the goals and summarize what they have just reasoned through. The multiple-choice questions force students to take a position and justify it; afterwards, the provided scaffolding helps students identify what was right and wrong in their thinking. Some of the written discussions contribute to this, allowing the students to compare and contrast multiple hypothetical student perspectives to decide which statements are correct. The tutorial thus implements innovation through induced struggle, followed by efficiency through scaffolding support and explanation during the "time for telling" [49] over many small steps throughout a single section.

These learning sequences build upon each other in each section, culminating in checkpoints that give a broad overview of the preceding sequences. These checkpoints serve as a break point for students to consider many related concepts at once, and to reflect on whether they make sense.

*Course implementation*

The tutorial was administered at a large research university in the United States in two separate courses. The first was the standard two-semester QM course for junior-/senior-level physics majors (for which calculus 1-3 and linear algebra are prerequisites), for which data were collected from two different implementations of the course in successive years by different instructors. Both instructors are physics education research-friendly and have used research-based learning tools before. There were 13 students in the first year and 22 students in the second year, for a total of 25 students. The tutorial was given by both instructors after the students had already learned the relevant foundational quantum mechanical concepts via traditional lecture-based instruction. The course text for this junior/senior level course was McIntyre's *Quantum Mechanics: A Paradigms Approach* in one year and Griffiths' *Introduction to Quantum Mechanics* in the other year, with additional material related to QISE covered in lecture by both instructors, including single-qubit and multi-qubit systems, single- and multi-qubit gates, quantum measurement, Bell measurement, quantum key distribution, and quantum teleportation. The instructor who used Griffiths's textbook started with the chapter on formalism (chapter 3), then covered spin-1/2 in chapter 4 before going back to chapters 1 and 2. Both instructors thus covered the material pertaining to two-state systems early in the course, before introducing wavefunctions and infinite-dimensional Hilbert spaces. The two instructors were very supportive of physics education research and have implemented physics education research-based pedagogies in their classes many times; in addition to this and other tutorials, they used a number of interactive-engagement Clicker Question Sequences in the course as well.

The second course was a multi-disciplinary undergraduate course titled "Foundations of Quantum Computing and Quantum Information." As the name suggests, the course focuses on an introduction to quantum information and (particularly) quantum computing. Aimed at undergraduate students from all science and engineering majors, it is the only mandatory course for a Quantum Computing and Quantum Information certificate available to interested undergraduate students across disciplines. The class comprised 28 students of diverse backgrounds: seven from engineering (engineering science, computer engineering, or industrial engineering); seven from computer science; six from math; two from chemistry; and twelve from physics. Some students chose more than one major (double major) from these disciplines. Most of them were sophomores, juniors, and seniors (one first-year student from the College of General Studies was enrolled). The structure of this particular course enables students of any background to benefit from taking it. The only prerequisites are Calculus 1 and 2, which serve as a proxy to determine students' ability to engage with the math, predominantly linear algebra, taught in the course in a self-contained way. The curriculum is generally focused on ideas such as two-state systems, quantum gates, and doing measurements, but is less concerned with the details of making a good qubit, correcting errors, and the technical physics behind gating and measurements. Selected as the course text was Wong's *Introduction to Classical and Quantum Computing*. In addition to the foundational QISE material covered in the QM course, the subjects of study included quantum algorithms, quantum circuits, quantum architecture, and activities using the Qiskit software development kit provided by IBM. While in this course Clicker Question Sequences were not used, the instructor did assign several tutorials on different topics, including this one on quantum computing basics.

The quantum computing tutorial was implemented as homework after students had traditional lecture-based instruction on the relevant concepts. The students were first given the pre-test to establish a baseline level of knowledge following the lectures and traditional homework, and then they were assigned the tutorial for homework. Afterwards, once students submitted the homework, they were given the post-test; both tests were assessed on paper at an in-person setting, and all students were given enough time to finish. Regarding student grades, the pre-test counted for completeness grade and the post-test counted for correctness grade for their quizzes. Common difficulties and performance improvements are discussed in the following sections. Because some of these questions were discovered during student interviews to be relatively easy for Physics students, one of the Physics instructors opted not to include these questions (marked in this text with an asterisk [*]) over time and length concerns.

Two researchers graded a fifth of the pre- and post-tests. After discussion, they converged on a rubric for which the inter-rater reliability was greater than 90%. Following this, one researcher graded the remaining pre- and post-tests. Each student response to an open-ended question was graded on a three-tiered scale of 0, 0.5, or 1 point, for each salient unit of a problem. This was the case for all but three questions, which were scored all-or-nothing out of 1 point. Appendix B summarizes the criteria for each point threshold and provides samples of responses that earned half credit.

In total, the data that are matched for the same students across the pre-test and post-test comprise 18 students in the QCQI course (hereby referred to as "QCQI students") and 35 students in the physics course (referred to as "Physics students") who completed the pre-test, tutorial, and post-test. The 18 QCQI students who engaged with the tutorial were a subset of the 28 students in the QCQI course mentioned earlier. Additional unmatched data exist for the remaining 10 students who took the post-test but not the pre-test. These data are presented in Appendix C and discussed separately. Also in Appendix C are the results for Physics students differentiated by class, though in the main text we describe the results for the two Physics classes in aggregate.

**Results**

Overall, both Physics students (those in the QM course for physics majors) and QCQI students (those in the Foundations of Quantum Computing and Quantum Information course) did very well on the post-test after engaging with the tutorial. In both courses, only a handful of questions remained somewhat difficult for students after the tutorial, and there were many for which they performed well even after lecture-based instruction. The data are shown in Tables 2 and 3. The post-test questions are reproduced in Appendix A. For additional insight discussed in later sections, Appendix C presents data for the QCQI students' performance split into physics majors and non-physics majors (Tables 8-9).

Some questions were omitted in some classes over concerns of time—namely, questions 9a-c, 10a-d, and 11 in one class of Physics students, and question 10d for the QCQI students. For the Physics data in Table 2, these questions are marked with an asterisk [*], and the class that forewent them was taught using McIntyre's text, though again we emphasize that the instructors of both classes employed a spins-first approach.

The data are interpreted using Hake's normalized gain $\left(g = \frac{post\% - pre\%}{100\% - pre\%}\right)$ [52], which calculates the fraction of maximum possible improvement that was achieved, and Cohen's $d$ $\left(d = \frac{post\% - pre\%}{\sqrt{\frac{1}{2}\left(\sigma_{pre}^2 + \sigma_{post}^2\right)}}\right)$ [53], which measures the average difference in units of standard deviation and indicates how large the effect is compared to other factors. Hake defined $g < 0.3$ as low, $0.3 \leq g < 0.7$ as medium, and $g \geq 0.7$ as high, and found that traditional lecture instruction resulted in an average $g$ of 0.23 while courses using interactive engagement had average $g$ 0.48. Cohen's guidelines interpret $d \approx 0.2$ as small, $d \approx 0.5$ as medium, and $d \approx 0.8$ as large effect sizes. Both measures possess strengths and weaknesses [54,55], and each can distinguish between cases that the other cannot. When these complementary measures are in agreement regarding the level of student improvement, we can be relatively confident that the effect is real—however, in some extreme cases discussed later where both pre-test and post-test scores are very close to ~90-100%, these two measures can lose value compared to the strength of the overall scores themselves. In cases where post-test score was lower than pre-test score, this was typically the result of little or no room for improvement, and we characterize such cases with normalized change $\left(\frac{post\%-pre\%}{pre\%}\right)$ described by Marx and Cummings [56].

**Table 2.** Results for Physics students ($N = 35$); questions with an asterisk [*] have $N = 22$ due to one instructor opting out of those questions over concerns of time for administration. Pre-test and post-test average scores, normalized (norm.) gain [52], and effect size as measured by Cohen's $d$ [53] are presented. Student data for Table 2 columns are matched. [†]In cases of $Pre \leq Post$, we follow the methodology for normalized change described by Marx and Cummings [56].

| Question | Pre (N=35) (*N=22) | Post (matched) (N=35) (*N=22) | Norm. Gain[†] | Effect size |
|---|---|---|---|---|
| 1a | 99% | 100% | 1.00 | 0.24 |
| 1b | 99% | 100% | 1.00 | 0.24 |
| 1c | 63% | 91% | 0.77 | 0.91 |
| 1d | 86% | 92% | 0.45 | 0.33 |
| 2a | 100% | 99% | -0.01 | -0.24 |
| 2b | 33% | 50% | 0.26 | 0.59 |
| 3a | 91% | 100% | 1.00 | 0.43 |
| 3b | 89% | 96% | 0.63 | 0.29 |
| 3c | 73% | 91% | 0.68 | 0.55 |
| 3d | 46% | 69% | 0.42 | 0.69 |
| 4a | 30% | 81% | 0.73 | 1.32 |
| 4b | 54% | 97% | 0.94 | 1.34 |
| 4c | 53% | 89% | 0.76 | 1.17 |
| 4d | 56% | 84% | 0.65 | 0.70 |
| 5 | 43% | 87% | 0.78 | 1.08 |
| 6 | 26% | 80% | 0.73 | 1.30 |
| 7a | 80% | 97% | 0.86 | 0.61 |
| 7b | 53% | 77% | 0.52 | 0.62 |
| 7c | 87% | 100% | 1.00 | 0.56 |
| 8a | 70% | 91% | 0.71 | 0.64 |
| 8b | 71% | 96% | 0.85 | 0.75 |
| 8c | 83% | 97% | 0.83 | 0.60 |
| 8d | 61% | 87% | 0.67 | 0.67 |
| *9a | 82% | 100% | 1.00 | 0.72 |
| *9b | 82% | 100% | 1.00 | 0.72 |
| *9c | 91% | 95% | 0.50 | 0.20 |
| *10a | 91% | 95% | 0.50 | 0.27 |
| *10b | 70% | 77% | 0.23 | 0.21 |
| *10c | 66% | 89% | 0.67 | 0.70 |
| *10d | 57% | 89% | 0.74 | 0.82 |
| *11 | 30% | 73% | 0.61 | 1.03 |

**Table 3.** Results for QCQI students ($N = 18$). Pre-test and post-test average scores, normalized (norm.) gain, and effect size as measured by Cohen's $d$ are presented. Student data for these four columns are matched. An additional column is included containing (unmatched) data from all students who completed the post-test ($N = 28$). (Question 10d was not asked of the QCQI students because of time concerns.) †In cases of $Pre \leq Post$, we follow the methodology for normalized change described by Marx and Cummings.

| Question | Pre (N=18) | Post (matched) (N=18) | Norm. Gain | Effect size | Post (unmatched) (N=28) |
|---|---|---|---|---|---|
| 1a | 100% | 97% | -0.03 | -0.33 | 98% |
| 1b | 100% | 96% | -0.04 | -0.32 | 97% |
| 1c | 58% | 89% | 0.73 | 0.63 | 82% |
| 1d | 86% | 93% | 0.50 | 0.21 | 93% |
| 2a | 100% | 100% | 0 | 0 | 100% |
| 2b | 50% | 67% | 0.33 | 0.35 | 55% |
| 3a | 100% | 94% | -0.06 | -0.33 | 93% |
| 3b | 86% | 100% | 1.00 | 0.34 | 96% |
| 3c | 78% | 94% | 0.75 | 0.39 | 93% |
| 3d | 47% | 69% | 0.42 | 0.49 | 64% |
| 4a | 50% | 86% | 0.72 | 0.70 | 89% |
| 4b | 61% | 89% | 0.71 | 0.55 | 89% |
| 4c | 56% | 92% | 0.81 | 0.76 | 87% |
| 4d | 67% | 92% | 0.75 | 0.51 | 86% |
| 5 | 50% | 81% | 0.61 | 0.53 | 86% |
| 6 | 39% | 83% | 0.73 | 0.78 | 82% |
| 7a | 78% | 94% | 0.75 | 0.34 | 95% |
| 7b | 47% | 78% | 0.58 | 0.58 | 75% |
| 7c | 69% | 100% | 1.00 | 0.64 | 100% |
| 8a | 86% | 92% | 0.40 | 0.13 | 95% |
| 8b | 89% | 100% | 1.00 | 0.30 | 98% |
| 8c | 81% | 90% | 0.50 | 0.24 | 90% |
| 8d | 58% | 89% | 0.73 | 0.57 | 82% |
| 9a | 89% | 100% | 1.00 | 0.28 | 96% |
| 9b | 89% | 100% | 1.00 | 0.28 | 100% |
| 9c | 92% | 94% | 0.33 | 0.07 | 96% |
| 10a | 72% | 100% | 1.00 | 0.66 | 98% |
| 10b | 61% | 67% | 0.14 | 0.11 | 71% |
| 10c | 47% | 74% | 0.50 | 0.52 | 67% |
| 11 | 25% | 64% | 0.52 | 0.70 | 46% |

*Difficulties assessed for all students*

*Questions with performance ≥ 85% after lecture-based instruction*

Students earned high scores of 85% or greater across the board for Questions 1a, 1b, 1d, 2a, 2b, and 3b, and 9c, suggesting that these lecture-based instruction was sufficient for these concepts.

Questions 1a and 1b asked students to identify all the possible independent states for two-bit and two-qubit systems. They overall did very well. For classical systems, the four states are the only ones available, while quantum systems are allowed to be in any linear superposition of the analogous four linearly independent basis states. Only in rare instances did students answer that both bits or qubits must match in the possible states of the system, thus yielding two possible states for their final answer. Out of all the students, this happened only for one Physics student on the pre-test and one QCQI student on the post-test. Another QCQI student stated that infinitely many states are available, such as $\{|00\rangle, |01\rangle, |+-\rangle, |++\rangle, ...\}$ (where the states $|\pm\rangle$ are defined to be $\frac{1}{\sqrt{2}}(|0\rangle \pm |1\rangle)$ for each qubit), so long as the states within the ket brackets are orthonormal. This was most likely connected to the fact that there are infinitely many bases that can be chosen for a quantum system. While this is true, the states are not all independent from one another, and so they cannot all be used as part of the response to the question. Interestingly, both QCQI students had answered the question correctly on the pre-test (see Table 3). Though exhibited rarely in these data, these are likely to be difficulties characteristic of students who may struggle with these concepts.

Questions 1d, and 9c as well, predominantly focused on using the Born rule to determine probabilities of measuring specific outcomes, for both individual and aggregate measurements. Students demonstrated little difficulty (see Tables 2-3) in taking the modulus squared of the coefficients associated with each basis state. Some common difficulties observed on Question 1d are discussed in the corresponding section ("*Probabilities of measuring each possible outcome*") below.

Additionally, students did well with Questions 2a and 3a. In each case, they were asked to provide the probabilities of obtaining certain outcomes when a qubit in a given state is measured (a specific arbitrary state for Question 2a, and the state $|-\rangle$ for Question 3a when measured in the $\{|+\rangle, |-\rangle\}$ basis). Students in both courses scored nearly perfectly on both questions on the pre-test.

Students also did reasonably well on Question 3b, which dealt with consecutive measurements made in a different basis; any mistakes made on the pre-test were minor and were overcome on the post-test (see Tables 2-3).

*Physical things that can be used as bits or qubits*

Question 1c asked students to give examples of physical entities that could be used to encode classical bits and qubits. Students in both courses improved on this question, mostly in their given examples of qubits. On the pre-test, some students did not provide answers for one or both parts of the question; of those who did, many students were able to give adequate examples of classical bits, though some responses like "a circuit," "an AND gate," or "voltage (on a wire)" were too vague and did not include the crucial element (e.g., an on-off switch for the circuit) that would have given their example two discernible states. On the pre-test, examples for qubits were somewhat less clear, including an issue of a similar nature (e.g., invoking electrons or photons very generally without specifying the two basis states that would make them suitable qubits; some students explicitly referred to the position of an electron). In addition to the preceding answers, some others appeared to refer to quantum systems for both answers, such as giving "ground and excited [states of a] hydrogen atom" or "a particle trapped in a box" as a classical bit.

In one or two cases on both the pre-test and post-test, students clearly described an idea that a classical bit is digital while a qubit is analogue, such as saying a pipe with water flowing or not can be taken as a classical bit, while the amount of water in the pipe would constitute a qubit. Another student gave the example of a switch (classical) and a slider (quantum). This may be an attempt to incorporate the knowledge that a qubit can be in any linear superposition of two basis states, in some sense being "between" the two states, but it does not take the quantum behavior of the system, including measurement collapse, into account. One of the experts consulted during the development of the tutorial had characterized a qubit using similar language with regard to this "analogue" nature, so depending on the context, this can be a useful way of describing the concept. Some students specifically drew the distinction that a qubit can be in a superposition or "combination of states," while a classical bit could only be in one state, which is a helpful direction even if they were not specific about the type of superposition.

A small number of students responded by saying that the same thing, such as the heads and tails sides of a coin, can be used for both. Some students were on the right track but gave somewhat incomplete answers for a qubit, such as "an electron" or "particle spin" (because only spin-1/2 particles such as electrons would constitute a two-state system). Some students gave somewhat circular answers, such as saying "a bit string" when asked for a classical bit, or restating the fact that classical computers use classical bits and quantum computers use quantum bits. Responses of all these types were found on both the pre-test and the post-test, but more students gave correct and much more uniform answers on the post-test (see Tables 2-3).

The rate of correct answers climbed from around 60% to ~90% on the post-test (see Tables 2-3). The vast majority of students suitably indicated the two-state nature of their provided examples of both classical bits and qubits on the post-test, even if their pre-test answers were more ambiguous.

*Probabilities of measuring each possible outcome*

Question 1d gave students a generic state $a_{00}|00\rangle + a_{01}|01\rangle + a_{10}|10\rangle + a_{11}|11\rangle$, asking them what the probabilities are of measuring each of the four possible outcomes ($|a_{ij}|^2$ for $i,j = 0,1$) and for the sum of these probabilities $\left(\sum |a_{ij}|^2 = 1\right)$. Since most students indicated that the probabilities should sum to 1, they were awarded a baseline of half credit for this question regardless of their answer to the other part. However, on the pre-test, many students said that a measurement would result in any of the four outcomes each with a ¼ probability, without obvious justification for why this should be so. Others, particularly the Physics students, used the $a_{ij}$ coefficients, but took the simple square instead of the square of the modulus. This may be a difference in what the courses emphasized; the physics students could have seen a preponderance of real-number-only examples in class examples and homework assignments, while the QCQI class may have used many complex numbers, kept things abstract, or constantly reinforced the importance of the absolute value brackets, etc. Starting from relatively high pre-test scores of 86% in both courses, students improved a bit more on the post-test (see Tables 2-3).

*Application of Hadamard gate to a state*

Question 3c asked students for the result of applying a Hadamard gate to the $|1\rangle$ state, and the possible outcomes and their associated measurement probabilities. Most students got this question fully correct on the pre-test, but the ones who didn't left it blank, answered only one of the two parts correctly, made little progress, or wrote the Hadamard gate correctly but mistakenly wrote the $|0\rangle$ state (as they were asked on the pre-test) as $\binom{0}{0}$ instead of $\binom{1}{0}$, which made them conclude that application of the gate to the null vector results in the null vector. This misinterpretation of the $|0\rangle$ as the null vector was observed in only one of the two groups of Physics students. Post-test performance on this question is high at over 90% (see Tables 2-3), though students were asked to apply the Hadamard gate to the state $|1\rangle$ instead of $|0\rangle$, and so they were not given the opportunity to demonstrate whether they had corrected this particular error.

*Measurement of the output of a quantum computer*

Related to the preceding discussions of measuring outcomes in Questions 1d, 2a, and 3a-c, in which students did reasonably well (73% and higher) on both the pre-test and post-test, Questions 7a and 7c apply these ideas to a quantum computation. The trend continues, with students achieving comparatively high scores of 69% and higher on the pre-test after lecture-based instruction, and further improvements nearing 100% on the post-test. The questions ask what the probability is of measuring the outcome associated with $|01\rangle$, assumed to be the "correct answer" within the context of the problem, with the given state being $\frac{1}{2}\big(|00\rangle + |01\rangle + |10\rangle + |11\rangle\big)$ in 7a and $\sqrt{\frac{1}{300}}|00\rangle + \sqrt{\frac{99}{100}}|01\rangle + \sqrt{\frac{1}{300}}|10\rangle + \sqrt{\frac{1}{300}}|11\rangle$ in 7c. Regarding most of the scores of 0 on the pre-test, a number of students wrote for 7a that the right and wrong answers each had a 50% chance of being measured. Interestingly, for Question 7c, students typically gave a blank answer instead of an incorrect one.

Question 7b gave the generic state $a_{00}|00\rangle + a_{01}|01\rangle + a_{10}|10\rangle + a_{11}|11\rangle$ and asked students what $a_{ij}$ must be for a measurement to output the correct answer $|01\rangle$. On the pre-test, a large number of students stated clearly that $a_{01}$ must be 1 while the remaining $a_{ij}$ must be 0. Other students, particularly in the Physics course, were not as clear in their answers, e.g., saying $a_{ij} = 1$ without specifying any particular $i$ or $j$; still other students in both courses appeared completely lost or left the question blank (see Table 2). Students were still awarded partial credit if they said that the desired state can be achieved by applying the correct quantum gates.

Question 7b is further discussed later in connection with other questions involving quantum gates. Students did much better on the post-test, with scores rising from around 50% to 75-80% (see Tables 2-3).

*N entities vs. $2^N$ entities*

Before engaging with the tutorial, after traditional lecture-based instruction, one common response was that a major difference between an $N$-bit classical and $N$-qubit quantum computer is that various things that are $N$ for a classical computer should be replaced with $2^N$ for a quantum computer (e.g., $2^N$ qubits must be initialized and $2^N$ bits of information are obtained as the output of the computation on the quantum computer). This type of reasoning primitive led many students to incorrectly think that there are only $N$ distinctly different states available when computation takes place on a classical computer. It also led many students to use as a heuristic the notion that the processing and storage advantage of quantum computers can be understood by replacing $N$ for a classical computer with $2^N$ for a quantum computer. Our research points to such replacement being a reasoning primitive that students use [57], due to overgeneralization of the core ideas that they are taught regarding quantum computers working with superpositions of $2^N$ states in superposition, and the exponential advantage that quantum computers can provide for certain problems. Such a reasoning primitive can go unexamined if the precise meaning of such statements as "quantum computers can be in many more states than classical computers" is not explored. Further research can clarify and elaborate on the nature of these ideas.

Even in our earlier investigations before the development of the tutorial, we found that students often had difficulty correctly ascribing the numbers $N$ and $2^N$ to various quantities for a classical and quantum computer with $N$ bits or qubits, respectively. Therefore, across the pre-test and post-test that accompany the tutorial, this distinction between the two numbers ($N$ vs. $2^N$) was asked across several questions on the pre-test and post-test in multiple forms: in terms of number of operations during initialization (Questions 4a and 6), number of bits of information extractable from the output (Question 4b), number of linearly independent parameters necessary to describe a quantum computer and whether this severely limits its size (Question 4c), and number of available states or basis states (Question 5). For all these questions except 4c, the cases of classical and quantum computers were juxtaposed with one another; Question 4c contained this juxtaposition implicitly, since obviously classical computers with $N > 300$ can be built with no issue. The core issue is the tendency of many students to think that an $N$-qubit quantum computer has more linearly independent states than a $N$-bit classical computer, likely the result of the reasoning primitive suggesting that $N$-qubit quantum computers are associated with $2^N$ qubits or states while classical computers are associated with $N$ bits or states. Research shows that this may also be due to the difficulty in distinguishing between linearly independent states vs. the infinitely many superposition states one can construct for $N$ qubits.

Pre-test performance on Questions 4a-c, all true/false questions with explanation, was low among the Physics students and somewhat higher among QCQI students (around 50% in all cases), but improved substantially to above 80% on the post-test in both courses (see Tables 2-3). In each case, students were observed to defend their incorrect answers, provide no explanation, or leave the questions blank entirely on the pre-test. On the post-test, they tended to answer with more clarity and correct explanations.

Question 4a was particularly difficult for students in both courses, with many students on the pre-test agreeing with the statement that quantum computers must initialize $2^N$ qubits while classical computers need to initialize only $N$ bits. This is consistent with their answers in Question 6, for which the correct number of operations that must be performed to initialize an $N$-qubit quantum computer is $N$; many students chose $2^N$ on the pre-test. On the post-test, students performed better on both questions (see Tables 2-3), frequently converging on the reasoning that a classical and quantum computer of equal size must both initialize the same number of qubits. Most of those students identified that number to be $N$, though a small number said $2^N$ instead. This was observed on both the pre-test and the post-test. As these two questions, which did not appear consecutively, were effectively asking the same thing, students' consistency in their answers is a good sign that they are understanding and reasoning through what they are being asked.

For Question 4b on the pre-test, many students did not agree that a quantum computer's output, the result of a quantum measurement, contains an amount of information equivalent to that of a classical $N$-bit string. Again, many students gravitated toward the idea that a quantum computer must in some way contend with more information than an equivalently sized classical computer. This seems a natural assumption if one is to think at face value that quantum computers can offer some advantage over classical computers. After engaging with the tutorial, most students on the post-test correctly noted that both classical and quantum computers' outputs contain the same amount of information (see Tables 2-3).

It appears that students develop the sense that an $N$-qubit quantum computer has more linearly independent states than an $N$-bit classical computer has states (both have $2^N$) because a quantum computer can evolve its qubits in various states of superposition or entanglement during a quantum computation. In this vein, Question 4c appeals to the notion that keeping track of the amplitudes of each of these basis states, a problem not encountered on classical computers, is an infeasible task as $N$ becomes large. On the pre-test, many students agreed with the incorrect statement that this issue prevents large quantum computers from being built. Encouragingly, on the post-test, most students provided valid answers (see Tables 2-3), often stating explicitly that a quantum computer "keeps track" of its own qubit states and does not need to store associated information in classical registers, thus acknowledging that operating a quantum computer is different from simulating a quantum computer on classical hardware.

Students also learn that an $N$-qubit quantum computer can be in a superposition of all $2^N$ basis states while an $N$-bit classical computer can only be in one of the possible states at any time. It may not be explicit to students that an $N$-bit classical computer has $2^N$ possible states, and there is often a temptation to assert that there are fewer. This is consistent with the reasoning primitive that a classical computer ought to have $N$ available states. We observe on the pre-test that very many students in both courses agreed with the student in Question 5 who says, incorrectly, that a classical computer has only $N$ states available to it. Some students defended Student 1's reasoning while others provided no further explanation. The correctness figure of this question rose from around 50% on the pre-test to 80% on the post-test (see Tables 2-3), with the vast majority of students correctly agreeing with Student 2, many again citing that the number of possible classical states and quantum basis states should be equal.

*Superposition vs. entanglement*

Question 4d investigates whether students can distinguish between superposition and entanglement. It is not uncommon for students to confuse the two, even though they describe distinct phenomena. Entanglement becomes salient when there are two or more qubits involved in a system of interest, but many students easily make the overgeneralization that any multi-qubit state that is a linear combination (i.e., superposition) of the possible basis states is an entangled state. Only multi-qubit states that cannot be written as a product of states of each qubit are entangled. For example, entanglement can make the outcomes of measurement of the qubits dependent on each other. To entangle qubits, the qubits in question must have interacted either directly or indirectly with one another beforehand, a concept that was challenging for many students.

The pre-test scores were low (~50%) for the Physics students and higher for the QCQI students. In both courses, students' incorrect pre-test responses tended to be ones without elaboration or blank, suggesting they were not quite sure of their answers. Those QCQI students who answered the question correctly on the pre-test supplied the reason that only states that are non-factorable (in any basis) as the product of states for each qubit are considered entangled, and that not every superposition state of multiple qubits fulfills this criterion. While there are many common valid justifications for disagreeing with the incorrect statement, this was the most common one, as it was likely to be the one most emphasized in class. The tutorial helped students reflect upon this type of explanation as well as others. Both Physics and QCQI students improved on the post-test (~80%), and some also noted that, for a multi-qubit system, the set of all superposition states is larger than the set of all entangled states.

Questions 8a-d also probe students' knowledge of these concepts. Question 8a asks students if a single qubit can be in a superposition state, Question 8b asks if a single qubit can be in an entangled state, and Question 8c asks if multiple qubits can be in an entangled state. The QCQI students did quite well on the pre-test for Questions 8a-c (~80% and higher, see Table 3), while the Physics students had somewhat lower performance (~70-80%, see Table 2). For example, a small number of Physics students (but no QCQI students) were prone to saying that a single qubit can be entangled, usually with no elaboration; one student took for granted that entanglement was possible, saying only that entanglement is present for some but not all states, which is true only for multi-qubit systems. The number of affirmative responses for 8b decreased on the post-test. In both courses, pre-test performance on Question 8d, which specifically asked if qubits can become entangled without any prior interaction, was lower than for 8a-c. Question 8d was intended to be slightly trickier than the rest (~60% on the pre-test), and a number of students stated that entangled qubits need not have prior interactions (direct or indirect), with some saying that simply passing them through an appropriate gate, or that preparing them together at the same time, would entangle them. These students appear to have learned about gates, e.g., control not (CNOT) that can entangle qubits, but had not yet made the connection that a gate like CNOT that entangles qubits must involve some sort of direct or indirect qubit interaction. It may point to a rather mechanical approach to learning these types of concepts, invoking specific gates without consideration, e.g., to what applying those gates to a multi-qubit system would physically imply. Also on the pre-test, a number of students did not provide elaboration, pointing to the likelihood that they were unsure of the reasoning behind their affirmative or negative answers. For all of these questions, all students in both courses did very well on the post-test, with scores around 90% and higher (see Tables 2-3).

*State of student knowledge on single-qubit quantum gates*

Question 2b asked students how a single-qubit quantum gate might be constructed to transform an arbitrary given state to one of the basis states. Similarly, Question 3d asked students to provide a quantum gate that would transform one basis state to the other basis state (for which the Pauli X gate $\sigma_x$ is an example). Since numerically solving for such a quantum gate can be quite mathematically involved, we considered qualitative responses specifying some sort of rotation of the input state perfectly acceptable. Students typically either gave this type of response associating the application of quantum gates to rotation of the state, or alternatively provided a gate represented by a matrix for such questions, which almost always were projective measurement gates such as $\begin{pmatrix} 1 & 0 \\ 0 & 0 \end{pmatrix}$ or one of the familiar Pauli gates. A few students even explicitly mentioned that the gate should collapse the given state to the desired outcome, not recognizing the significance of the unitarity of quantum gates. Some gave gates like NOT or CNOT which were either classical or multi-qubit in nature. Other students simply responded to Question 3d by affirming that such a gate does exist, without providing further elaboration or the gate itself.

It was interesting that, for both questions, there was a distinct split between courses in how students tended to answer this question. Only about half of QCQI students, but nearly all Physics students, gave a matrix as an answer, and some students in both courses even acknowledged that their chosen matrix could not be a valid quantum gate because it was not unitary or did not preserve state normalization. This is most likely due to a difference in instruction; the question asked, with intentionally broad phrasing, "how might you construct a quantum gate…" The QCQI students may have focused on *how* to construct a gate and what such gates accomplish while Physics students may have taken the question to be asking them to construct and present an appropriate gate in matrix notation. Given that much of the Physics coursework involves matrix manipulations, they may have viewed this to be the most suitable course of action based upon the question prompt.

For Question 2b (describing a gate that transforms a given state to a basis state), some students, upon seeing that their projective gate $\begin{pmatrix} 1 & 0 \\ 0 & 0 \end{pmatrix}$, when applied to the given state $\sqrt{\frac{11}{13}}|0\rangle + \sqrt{\frac{2}{13}}|1\rangle$, resulted in the state $\sqrt{\frac{11}{13}}|0\rangle$, attempted to achieve a normalized state by multiplying their provided gate by $\sqrt{\frac{13}{11}}$. While our grading rubric did not give such responses full credit (see Appendix B), we see that they are applying the rules they learned to situations not explicitly discussed in the tutorial in the best ways that they know how. A few students gave exemplary responses for this question by specifying that the gate must be unitary and providing the additional constraints that render the problem soluble. Such students received full marks regardless of whether they were able to numerically complete the calculation to arrive at a suitable quantum gate. About one or two students from each class provided such answers for both Question 2b and 3d; while the remaining students did not articulate such concepts with the same precision, many QCQI students did make some mention of a "rotation," referring to the fact that quantum gates are rotations of quantum states in a Hilbert space. This was likely to have been emphasized more in the QCQI course than the Physics course.

An additional point can be made about Question 7b, a two-part question that asks students what the probability amplitudes of a state $a_{00}|00\rangle + a_{01}|01\rangle + a_{10}|10\rangle + a_{11}|11\rangle$ must be if one desires to read an output of 01, and how these probability amplitudes can be achieved if the initial state of the quantum computer did not have them. Most students successfully provided the probability amplitudes but did not answer the second part on the pre-test; a few who did answer said that this was impossible. The expected answer is that quantum gates can be applied to change the state of the system from an initial to a final state. On the post-test, there was substantial improvement in both completion and correctness on this question (from ~50% to ~80%, see Tables 2-3), with most students citing quantum gates in some way. A few instead suggested, apparently assuming an infinite ensemble of quantum computers in superpositions of all four basis states, something to the effect of continuously sampling from the ensemble until one obtained the desired state as the output. Though these methods in some respects defeat the purpose of running a quantum computation, they were accepted as valid answers.

*Difficulties assessed with a subset of students (excluding one class of Physics students)*

These questions were asked of students in the QCQI class and one of the Physics classes (using Griffiths's text); the other Physics class (using McIntyre's) eschewed them over concerns of time. (We mention the textbook correspondences for transparency, but we consider both Physics courses to be taught in spins-first fashion with not significant differences in how relevant topics were introduced in lectures.)

*Outcomes of measurement*

Students did relatively well, over 80%, on Questions 9a-c on the pre-test, which asked them to provide the possible outcomes of various hypothetical measurements on given states (these questions were not posed to one of the Physics classes' pre-test and post-test, as noted in Table 2). They improved to nearly 100% on the post-test (see Tables 2-3). These questions were similar to 2a and 3a, but with consecutive measurements. The most common mistakes involved not recognizing that the state would collapse after measurements.

*Finding the bra state corresponding to the given ket state*

On the pre-test for Question 10a, which asked for the corresponding bra state to the given ket state, the Physics students performed excellently (91%, see Table 2) and QCQI students did reasonably well (72%, see Table 3), and both achieved near 100% on the post-test. Splitting the QCQI students between physics and non-physics majors reveals that the physics students were overall more comfortable with bra states on the pre-test, likely because many of them had already taken QM (see Appendix C). The most common mistake was not finding the complex conjugate of the expansion coefficients. Some students left this question blank.

*Difficulties with outer products, operators, and Dirac notation*

For pre-test Question 10b, which gave students an operator $a|0\rangle\langle 0| + b|0\rangle\langle 1| + c|1\rangle\langle 0| + d|1\rangle\langle 1|$ and asked them to fill in the values of $a, b, c, d$ for the identity operator, most students were unable to do so correctly. For pre-test Question 10c, many students were unable to provide a fully correct outer product. For both questions, many of the student responses gave the same value for all four matrix elements, which is likely to be a guess in the absence of better options. This could be connected to many students' tendency to say that all four outcomes are equiprobable in Question 1d. Others left the question blank. Given this apparent lack of fluency, it is unclear also whether the students would have had an easier time had they been asked to write their answer in matrix representation instead. In a typical physics course, students have been found to have difficulties distinguishing an inner product from an outer product, but this is almost universally corrected after engagement with a research-based learning tool, like a clicker question sequence or a tutorial [45]. On a related note, Question 10c asked students to find the outer product $|q\rangle\langle q|$; some students stated that the identity operator was $\hat{I} = |q\rangle\langle q|$, which appears to be conflating it with the spectral decomposition $\hat{I} = \sum |q_i\rangle\langle q_i|$ where $i$ enumerates each linearly independent basis state.

For Question 10c, which was similar to Question 10b but asked students to find the values of $a, b, c, d$ for the outer product $|q\rangle\langle q|$ for a given $|q\rangle$, there was some major improvement on the post-test of roughly 30% in both classes. Question 10b, on the other hand, saw little change, despite the available room for improvement (see Tables 3-4). It is possible that students did not have much practice with outer products after being exposed to them.

Question 10d, asked of only one class of Physics students, indicates that calculating the outer product did not come particularly naturally after traditional lecture-based instruction, with improvement on the post-test after the tutorial (57% to 89%, see Table 3). Some students were clever enough to check their answers to the expression for $|q\rangle\langle q|1\rangle$ by recognizing that $\langle q|1\rangle$, an inner product, is simply a coefficient by which the state $|q\rangle$ is multiplied. Quantum physics students' comfortability with inner products compared to outer products has been documented in the past [45].

Question 11 asked students to write the Hadamard gate in Dirac notation. Students in the QCQI course who were physics majors may have had less difficulty if they had already taken QM, but it appeared to be quite difficult overall. Though some students answered the question correctly at least once, many left the question blank on the pre-test, or admitted to not knowing what the question was asking, and continued to struggle on the post-test. Answers on both the pre-test and post-test varied, and included ket states like $\frac{1}{\sqrt{2}}(|0\rangle - |1\rangle)$, which may indicate that these students were not comfortable with the distinction between ket states and outer products. Some students provided the correct answer in the form $\frac{1}{\sqrt{2}}(|0\rangle\langle 0| + |0\rangle\langle 1| + |1\rangle\langle 0| - |1\rangle\langle 1|)$, but others responded with $|0\rangle\langle +| + |1\rangle\langle -|$ or $|+\rangle\langle 0| + |-\rangle\langle 1|$, which were not used in the QCQI course text and appeared to be a notable instance of student creativity. Overall, there was some improvement from the pre-test (~30%) to the post-test (~70%; see Tables 3-4), but the quality of answers was not uniform.

A summary of the difficulties discussed in this section can be found in Table 4.

**Table 4.** Summary of student difficulties as well as students' improvement on these difficulties between the pre-test and post-test. Difficulties and associated questions that were assessed with only a subset of the students are marked with an asterisk (*). In this table, "High pre-test performance" refers to that of 80% or higher; "major improvement" indicates score differentials of around 30%, normalized gains > 0.60 and effect sizes > 0.50, and "some improvement" indicates improvements with normalized gains > 0.30 and effect sizes > 0.20 (only Question 10b does not meet either threshold).

| Difficulties | Pre-/post-test # | Comments |
| --- | --- | --- |
| Enumeration of all possible states or linearly independent states for a system of two bits or qubits, respectively | 1a, 1b | High pre-test performance |
| Specifying outcomes of measurements made on qubits, including consecutive measurements made in the same or a different basis | 2a, 3a, 3b, 3c, 9a*, 9b*, 9c* | High pre-test performance |
| Correct application of the Hadamard gate (given to students) to specified single-qubit states | 3c | High pre-test performance |
| Inability to provide a physical example of a qubit | 1c | Major improvement |
| Probabilities of measuring each possible outcome | 1d | Some improvement; high pre-test performance |
| Difficulties with reasoning about various possible states of a quantum computer | 7a, 7b, 7c | Some improvement |
| Difficulties with $N$ vs. $2^N$ entities | 4a, 4b, 4c, 5, 6 | Major improvement |
| Conflation of superposition and entanglement | 4d, 8a, 8b, 8c, 8d | Major improvement in all cases except in cases of high pre-test performance (QCQI 8a, 8b, 8c; Physics 8c) |
| Difficulties with what single-qubit quantum gates accomplish | 2b, 3d, 7b | Some improvement |
| *Finding the bra state corresponding to the given ket state | 10a* | Some improvement |
| *Difficulties with outer products and operators | 10c* | Some improvement |
| *Difficulties with expressing quantum gates in Dirac notation | 11* | Major improvement (QCQI has normalized gain 0.52, but pre-test figure is particularly low at 25%) |

*Long-term learning gains at the end of the semester*

In addition to the pre- and post-test data, we also collected data on longer-term learning retention for some questions for one of the physics classes ($N = 22$). Details can be found in Table 4 in Appendix B. Students were evaluated on the concepts covered in Questions 4a-d, 5, and 10b on the final exam, which was given several months after the material was covered and the tutorial was assigned. All of these questions dealt with concepts regarding quantum computing except for 10b, which asked them to enumerate the matrix elements of the identity operator. Overall, students did very well, scoring no worse than 73% and nearly 100% for four of the questions. Question 5 asked about the number of states (for classical computers) or linearly independent states (for quantum computers) available during a calculation; the somewhat lower performance here could be the result of the question presenting students with only one statement to evaluate rather than two statements to compare, which could possibly reduce the salience of the reasoning primitive that a classical computer has $N$ available states compared to a quantum computer's $2^N$ states.

Only a marginal improvement on 10b indicates that the identity operator given as an outer product and identification of suitable coefficients continues to be challenging for some students, with the most popular incorrect answer of all matrix elements being equal to one another still noticeably present. This seems to be a strong alternative conception when this question is framed in terms of outer products. Prior investigation shows that a comparable percentage of Physics students struggled when prompted to write the identity operator as an outer product of a complete set of orthonormal states, even on the post-test. Though Question 10b instead provided the generic outer product for which students were to identify the correct coefficients *a-d*, this indicates that students can routinely have difficulties with the identity operator expressed in Dirac notation [41]. On the final exam, two students multiplied the operator by an overall factor as though normalizing a ket state. This is curious, as it was observed only twice throughout the pre-test and post-test across all classes ($N = 63$), so it appeared with higher frequency on the final ($N = 22$). Additionally, one student wrote the identity operator in matrix form for a four-dimensional (instead of two-dimensional) Hilbert space. Some of the other students who noted that all four coefficients should be 1 may have also been thinking about an identity matrix in four-dimensional Hilbert space, possibly assuming that they were giving coefficients for only diagonal elements as opposed to coefficients for a $2 \times 2$ matrix with both diagonal and off-diagonal elements.

**Conclusions and discussion**

*Largest differences between QCQI and Physics students*

It is somewhat surprising that the QCQI students performed better on many pre-test questions than Physics students, though in retrospect there is sense to this. Our initial expectations were that Physics students might have a better grasp of the underlying QM formalism and thus may be able find many salient connections to the material. However, we conclude that the QCQI course focused on those aspects in much more depth with no distraction from other types of course material than a typical quantum physics course for physics juniors/seniors would, even though both physics instructors teaching the latter course noted that they had covered the materials via lecture-based instruction. In particular, the QCQI students did better overall possibly because they had focused on this type of material (e.g., on qubits, multi-qubit systems, gates) for the entire semester and the topic of basics of quantum computing was not an addition to the typical material covered as in the junior/senior level Physics course.

Questions 2b, 3d, 4a, 4d, and 6 are ones for which the post-test performance differed by more than 15% between the Physics and QCQI students, with QCQI students scoring higher for all cases (see Tables 2-3). Questions 2b and 3d both dealt with conceptual understanding of quantum gates, giving students a starting state and asking them to comment on possible gates that could be applied to reach a desired state. Physics students were more likely to explicitly come up with a gate to apply to the state while QCQI students provided more conceptual statements. Since calculating a valid gate was a complicated process, and most students' numerical responses in both courses did not succeed at doing so, this may have put the Physics students at a disadvantage. Questions 4a and 6 asked students whether particular enumerations in classical and quantum computers were $N$ or $2^N$, which was another difficult concept for both populations on the pre-test, but one on which the QCQI students likely spent much more extensive class time. Question 4d asked students to explain the difference, if any, between superposition states and entangled states, and once again, this was a key concept that the QCQI course focused on.

In the data shown in Table 3, the QCQI students' performance decreased by more than 10% when considering all students (unmatched data) for Questions 2b and 11. Given that the unmatched data comprise an additional 1 physics major and 9 non-physics majors, it follows that the vast majority of this decrease is contributed by non-physics majors. Question 11 dealt with Dirac notation, so non-physics majors in the course may have had more difficulty, or less prior knowledge, concerning quantum gates (Question 2b) and Dirac notation, particularly as it concerns outer products, which was the most difficult concept for students. This is investigated further in the following section, "*Comparison between physics and non-physics majors in the QCQI course*."

With all this said, both QCQI and Physics students improved on these questions from the pre-test to the post-test. However, it was a trend in both courses that if the starting score was too low, about 50% or under, it ultimately did not rise to encompass the vast majority of the class as was observed in some other cases—such questions with low pre-test scores saw an improvement to roughly half correctness on the post-test. This is in line with previous observations that these research-based tools work best if there is enough knowledge in the class already, which can be a combination of students' preparation, what they were able to learn from lecture, or other factors [58]. When students were answering questions on the pre-test with roughly half correctness, they were almost always able to improve to near full correctness after they engaged with the tutorial.

*Comparison between physics and non-physics majors in the QCQI course*

A bit of additional insight may be derived by comparing, within the QCQI course, the post-test performance of the physics majors to that of the non-physics majors. These results can be found in Appendix C.

When comparing the matched data, discrepancies of 15% or greater are observed in Questions 1c, 2b, 10c and 11. When considering all of the students through the unmatched data, discrepancies of 15% or greater are observed in Questions 2b, 3a, 4c, 8d, 10c, and 11. Questions 2b, 10c and 11 are common to both the large and small datasets, with physics majors doing better, so in general it appears that gates, outer products, and Dirac notation are concepts that physics majors have an easier time with than non-physics majors; these findings remain tentative in light of the small size of the datasets, and caution should be taken regarding their generalizability. It is worth noting that roughly half of the physics majors in this course had completed the first semester of their upper-level QM course, which emphasizes these topics in the curriculum. Physics majors also did slightly better on Questions 4c (a question targeting the incorrect notion that the limit on building large quantum computers is the inability to contain the data that quantum computers handle) and 8d (a question on entanglement). It is plausible that, though these concepts are not as widely taught in a physics-focused QM course, physics majors' performance could still be related to familiarity with and time spent on related ideas in QM. Likewise, Dirac notation was taught in both courses, and the Physics students did modestly but not necessarily conclusively better than the QCQI students on the pre-test and post-test.

Some differences in post-test performance can also be seen between the physics and non-physics majors with Questions 1c and 3a in either the matched and unmatched data. As is the case with Questions 4c and 8d, however, the performance of both groups was high, and the difference amounts to a total of roughly two students performing better on the part of one group or the other.

Some differences were also observed between matched and unmatched non-physics major data, with drops of more than 10% for Questions 1c, 4d, 8d and 11 once all the non-physics majors are considered. This suggests that the non-physics majors as a whole are less comfortable than physics majors with physical examples of qubits, the difference between superposition and entanglement, and Dirac notation.

*Innovation and efficiency in the tutorial*

As discussed, one of the theoretical underpinnings informing the development of this tutorial is the innovation and efficiency framework. In one interpretation, innovation can be realized by presenting students with invention tasks and challenging questions to ponder, while efficiency can be conveyed by helping students engage with more highly scaffolded learning material; these phases can be incorporated together within a single guided learning sequence. The tutorial follows this structure by deliberately asking students questions that they will struggle productively with before providing more resolution through scaffolding, which can take the form of written explanatory passages or simulated hypothetical student discussions that position various common concepts against each other. We saw this progression play out in the interview stage, indicating that innovation and efficiency were balanced enough to prevent the "frustrated novice" phenomenon that would result from inducing too much innovative struggle. The varied tasks and large scope of breadth and depth of the tutorial also help with the development of adaptive rather than routine experts, and the overall high performance of students in both the interviews and course implementations suggest a successful fulfillment of that goal.

*Overall trends and instructional implications*

We found that the extent of student struggle with the material fell into several big categories. Some concepts were more challenging than others after traditional lecture-based instruction, with performance of less than 70% on the pre-test in both courses. These concepts included contending with the numbers $N$ and $2^N$ in the contexts of classical and quantum computers, e.g., in terms of how many operations must be performed to initialize each type of computer (seen in Questions 4a-c, 5, and 6) due to the reasoning primitive discussed. In particular, we identified a reasoning primitive [57] many students use for why a quantum computer may have a processing and storage advantage: these advantages come from the fact that certain entities for an $N$-bit classical computer get replaced with $2^N$ for a quantum computer. This reasoning primitive manifested in many key concepts, leading many students to incorrectly conclude that a $N$-bit classical computer only has $N$ available states for computation while an $N$-qubit quantum computer requires $2^N$ entities to be initialized at the beginning of the computation, or that an $N$-qubit quantum computer works with $2^N$ bits of information during the quantum computation or when yielding its output after measurement. Another challenging theme was the nature and role of quantum gates, including the importance of unitarity and how they are used to manipulate qubits into a desired state (seen in Questions 2b, 3e, and 7b). Students also experienced challenges with the difference between superposition and entanglement (seen in Questions 4d and 8d). Students also initially had some difficulty providing physical examples of qubits (seen in Question 1c), though most of these were a matter of not specifying the two basis states of the qubit (while most did specify two states for their examples of a classical bit). Dirac notation (Questions 10b, 10c, and 11) also tended to be a difficult topic at the outset, and in particular there was little to no improvement for Question 10b in either course.

Students performed well on the vast majority of the post-test questions, including for most of the concepts that they had initially struggled with. In this regard, the only questions whose scores remained under 70% on the post-test were the ones dealing with quantum gates (Questions 2b and 3c). In these cases, students typically knew very well how to find a matrix that would give them the desired output state, but these matrices tended not to be unitary, signaling that they had not yet internalized the criteria for a unitary gate regardless of whether their course had a Physics or QCQI focus. Additionally, for the students in the QCQI course, post-test performances of under 70% were also observed when writing outer products in Dirac notation to make up a quantum gate (Questions 10b and 11). The QCQI students, especially non-physics majors, may have had some more difficulty with operators in general; it should be noted, however, that while the Physics students did clear the (arbitrary) 70%, mark on those questions, it was not by any large margins. The QCQI students also had slight trouble with deciding the probabilities of getting "right" and "wrong" answers from a quantum computation (Question 7c).

Finally, students in both courses tended not to struggle with providing the possible joint states of two bits or two qubits (Question 1a-1b), any questions involving measurement outcomes and probabilities of qubits that may or may not involve measurement collapse (Questions 1d, 2a, 3a-3c, 7a, 9a-9c), and situations involving complex conjugates (Questions 1d and 10a). These concepts are well-conveyed by traditional lecture instruction.

Despite the QCQI students' greater variance in background preparation, the post-test scores were quite comparable in the two courses after engaging with the tutorial; there were remarkably few differences between the QCQI and Physics students on the whole. In light of these findings, we suggest that students in both courses could benefit greatly from more discussions, considerations, and improvement of instructional design regarding (1) the mechanics and conceptual function of quantum gates, and (2) the appropriate attributions of $N$ and $2^N$ for various quantities, which includes, e.g., discussions that a quantum computer's "exponential advantage" does not at all imply that a classical computer has only $N$ instead of $2^N$ possible states. It would especially (but not exclusively) benefit QCQI students to provide more guidance and scaffolding on (3) Dirac notation and (4) outer products; unlike the Physics students, they had not had extensive prior practice with such concepts, which have been documented to be challenging in their own right even in Physics contexts [41,45].

Most of the other concepts covered by the tutorial seem to be handled rather well by both QCQI and Physics students, so additional instruction may be less necessary for those concepts compared to the ones cited here. The tutorial itself generally reinforced high pre-test scores while substantially boosting lower ones and thus appears to be beneficial for students in both courses. This suggests that, even with different instructors and different course contexts and broader goals, a research-based tutorial can be useful in helping students learn the basics of quantum computing. With the need for effective instruction to help students understand and communicate challenging foundational concepts in QISE, such a tutorial can play a key role in fulfilling this promise.

## Acknowledgments

We thank the NSF for award PHY-2309260. We thank all students whose data were analyzed and Dr. Robert P. Devaty for his constructive feedback on the manuscript.

**Appendix A**

The questions that students were given for the post-test are reproduced here. All parts of Questions 1-8 were given to students in the QCQI class and both Physics classes. All parts of Questions 9-11 were given to students in the QCQI class and only one of the Physics classes (with the exception of 10d, which was given to only the Physics class). Coding details for student responses are provided in Appendix B.

Students were given the following information:

- All states, operators, and measurements are assumed to be in the $\{|0\rangle, |1\rangle\}$ orthonormal basis. In this notation, the states $|\pm\rangle = \frac{1}{\sqrt{2}}(|0\rangle \pm |1\rangle)$ for each qubit.
- A measurement made in immediate succession means that the measurement is to be made before the state has had time to evolve.
- The symbol $\doteq$ means "is represented by" in a given basis.

The pre-test differs from the post-test in the following questions, all of which are small changes made to the given states:

- Questions 2 and 9: $|q\rangle = \sqrt{\frac{13}{17}}|0\rangle + \sqrt{\frac{4}{17}}|1\rangle$.
- Question 2a: What is the probability of obtaining the state $|1\rangle$?
- Question 3: Consider the state $|q\rangle = |0\rangle$.
- Question 3a: What is the probability of obtaining the state $|+\rangle$?
- Question 3b: Suppose the measurement in part (A) yielded the state $|+\rangle$.
- Question 3c: Suppose we first act upon the state $|q\rangle = |0\rangle$ with the Hadamard gate, $\hat{H} \doteq \frac{1}{\sqrt{2}}\begin{pmatrix} 1 & 1 \\ 1 & -1 \end{pmatrix}$. What is the state after the gate has been applied, $\hat{H}|q\rangle$? Is it possible for a measurement in the state $\hat{H}|q\rangle$ to yield $|1\rangle$, and if so, with what probability?
- Question 3d: Can you construct a quantum gate $\hat{G}$ such that a measurement on the state $\hat{G}|q\rangle$ will then yield the state $|1\rangle$ with 100% certainty?
- Question 10: Consider the state $|q\rangle = \frac{1}{\sqrt{2}}(|0\rangle + i|1\rangle)$.

1. This question is about multiple bits and qubits.

    (A) How many total states are possible for a system of 2 classical bits? Write them down (e.g., "00" can represent the state in which both bits take values of 0).

    (B) How many total linearly independent states are possible for a system of 2 quantum bits (qubits)? Write down a representation of these states using the notation of kets, e.g., $|00\rangle$, ...

    (C) Give an example of a physical system that can be used as a classical bit, and one that can be used as a qubit.

    (D) The quantum state of the system of two qubits can be represented as $a_{00}|00\rangle + a_{01}|01\rangle + a_{10}|10\rangle + a_{11}|11\rangle$. If we make a measurement on this system, what are the probabilities of obtaining each of the basis states $|00\rangle$, $|01\rangle$, $|10\rangle$, and $|11\rangle$? What should you obtain by adding these 4 probabilities together? (Note: In general, assume complex numbers $\{a_{ij}\}$ where $i,j = 0,1$)

2. Consider the state $|q\rangle = \sqrt{\frac{11}{13}}|0\rangle + \sqrt{\frac{2}{13}}|1\rangle$.

(A) If we perform a measurement on a system prepared in the above state $|q\rangle$, what is the probability of obtaining the state $|0\rangle$? What is the probability of obtaining the state $|1\rangle$?

(B) How might you construct a gate such that, after its application to the state $|q\rangle$, a measurement will yield the state $|0\rangle$?

3. Consider the state $|q\rangle = |1\rangle$. In the $S_z$ basis, $|1\rangle \doteq \begin{pmatrix} 0 \\ 1 \end{pmatrix}$

   (Note: In terms of the $|0\rangle$ and $|1\rangle$ states, the states $|+\rangle$ and $|-\rangle$ are defined as $|+\rangle = \frac{1}{\sqrt{2}}(|0\rangle + |1\rangle)$ and $|-\rangle = \frac{1}{\sqrt{2}}(|0\rangle - |1\rangle)$.)

   (A) If a qubit in the state $|q\rangle$ is measured in the $\{|+\rangle, |-\rangle\}$ basis, what is the probability of obtaining the state $|-\rangle$?

   (B) Suppose the measurement in part (A) yielded the state $|-\rangle$. If we make a successive measurement in the $\{|0\rangle, |1\rangle\}$ basis, what are the possible outcomes, and the probabilities of those outcomes?

   (C) Suppose we first act upon the state $|q\rangle = |1\rangle$ with the Hadamard gate, $\hat{H} \doteq \frac{1}{\sqrt{2}} \begin{pmatrix} 1 & 1 \\ 1 & -1 \end{pmatrix}$. What is the state after the gate has been applied, $\hat{H}|q\rangle$? Is it possible for a measurement in the state $\hat{H}|q\rangle$ to yield $|0\rangle$, and if so, with what probability?

   (D) Can you construct a quantum gate $\hat{G}$ such that a measurement on the state $\hat{G}|q\rangle$ will then yield the state $|0\rangle$ with 100% certainty?

4. Indicate whether the following statements are true or false, and explain your reasoning in each case:

   (A) To initialize an $N$-bit classical computer, one must set $N$ bit values to zero. To initialize an $N$-qubit quantum computer, one must set $2^N$ qubit states to $|0\rangle$.

   (B) When an $N$-qubit quantum computer completes a computation and we "read" the output in the $\{|0\rangle, |1\rangle\}$ basis for each qubit, we obtain information equivalent to that contained in $N$ classical bits.

   (C) Large quantum computers, e.g., with 1000 qubits, cannot be made because they cannot possibly process $2^{1000} \cong 10^{300}$ variables, which is more than the number of atoms in the known universe.

   (D) If the quantum state of two qubits is in a superposition state in the basis consisting of $\{|00\rangle, |01\rangle, |10\rangle, |11\rangle\}$, then the state is said to be entangled.

5. Choose one of the following two statements and explain why you agree with it:

   **Statement 1:** In an $N$-bit classical computer, the computer has $N$ available states during the calculation, but in an $N$-qubit quantum computer, there are $2^N$ linearly independent states.

   **Statement 2:** In an $N$-bit classical computer, the computer has $2^N$ available states during the calculation, and in an $N$-qubit quantum computer, there are $2^N$ linearly independent states.

6. How many operations must be performed to initialize a quantum computer with $N$ qubits?
   a. $N$
   b. $2N$
   c. $2^N$
   d. None of the above

7. Quantum algorithms do not always give the correct answer with 100% probability. Suppose you have a quantum algorithm for a two-qubit quantum computer in which the final state is $|q\rangle = a_{00}|00\rangle + a_{01}|01\rangle + a_{10}|10\rangle + a_{11}|11\rangle$, and the correct answer (verified through some classical methods) corresponds to 01 (the corresponding state is $|01\rangle$).

   (A) Suppose the initial state of the quantum computer is $\frac{1}{2}(|00\rangle + |01\rangle + |10\rangle + |11\rangle)$, i.e., $a_{ij} = \frac{1}{2}$, $i,j = 0,1$. What is the probability that the quantum computer will give the correct answer? What is the probability that the quantum computer will give the wrong answer?

   (B) What must $\{a_{ij}\}$ be for a measurement to yield the correct answer $|01\rangle$ 100% of the time? In a quantum computer, how can $\{a_{ij}\}$ be made to take these values if the initial state is the one in part (A)?

   (C) If the state of the quantum computer is made to be $\sqrt{\frac{1}{300}}|00\rangle + \sqrt{\frac{99}{100}}|01\rangle + \sqrt{\frac{1}{300}}|10\rangle + \sqrt{\frac{1}{300}}|11\rangle$, what is the probability that the quantum computer will give the correct answer? What is the probability that the quantum computer will give the wrong answer?

8. Answer the following questions:

   (A) Can a single qubit be in a superposition state? Explain.

   (B) Can a single qubit state be an entangled state? Explain.

(C) Can a multi-qubit state be an entangled state? Explain.

(D) Can a multi-qubit state be an entangled state without any prior interactions between the qubits (either directly between two qubits or via other qubits)? Explain.

*The questions below were not asked in physics class 1.*

9. Consider the state $|q\rangle = \sqrt{\frac{11}{13}}|0\rangle + \sqrt{\frac{2}{13}}|1\rangle$.

   (A) If the result of a first measurement made in this state was $|0\rangle$, and another measurement is made in immediate succession before the state has had time to evolve, what is the probability of obtaining the state $|0\rangle$? What is the probability of obtaining the state $|1\rangle$?

   (B) If instead the result of the first measurement was $|1\rangle$, and another measurement is made in immediate succession, what is the probability of obtaining the state $|0\rangle$? What is the probability of obtaining the state $|1\rangle$?

   (C) If we perform a measurement on a large number of systems, each prepared in the above state $|q\rangle$, what fraction of those measurements will yield the state $|0\rangle$? What fraction will yield the state $|1\rangle$?

10. Consider the state $|q\rangle = \frac{1}{\sqrt{2}}(|0\rangle - i|1\rangle)$.

    (A) What is the corresponding bra state for this ket state?

    (B) A general operator in a two-dimensional vector space can be written in the form $a|0\rangle\langle 0| + b|0\rangle\langle 1| + c|1\rangle\langle 0| + d|1\rangle\langle 1|$. Find the values of $a, b, c,$ and $d$ that the identity operator $\hat{I}$ takes in this 2-D vector space.

    (C) A general operator can be written in the form $a|0\rangle\langle 0| + b|0\rangle\langle 1| + c|1\rangle\langle 0| + d|1\rangle\langle 1|$. Find the values of $a, b, c,$ and $d$ that the outer product $|q\rangle\langle q|$ takes.

    (D) Calculate $|q\rangle\langle q|1\rangle$. (This is the action of the operator $|q\rangle\langle q|$ on the state $|1\rangle$.)
    *This question was not asked in the QCQI class.*

11. Write an expression for the Hadamard gate using Dirac notation.

## Appendix B

A summary of the rubric using which student responses were coded is provided in Table 5. The pre-test and post-test questions can be found in Appendix A.

**Table 5.** A summary of the rubric using which student responses were coded for pre-test and post-test questions. Bulleted items represent distinct observed responses. (Questions 3a and 7 are included for completeness but were graded on a two-tiered scale.)

| Question | 0 (includes blank responses) | Half credit (0.5; all "OR" statements are exclusive) | Full credit (1) |
| --- | --- | --- | --- |
| 1a | Response is irrelevant (None observed) | Provides some states (e.g., only 00 and 11) | Provides all 4 states |
| 1b | Response is irrelevant (None observed) | Provides some states (e.g., only $|00\rangle$ and $|11\rangle$) | Provides all 4 states |
| 1c | Improper examples of bit and qubit; circular reasoning (e.g., "a bit string can serve as a bit") | Correct example of either bit OR qubit (must specify two-state entities); partial credit includes quantum systems with > 2 states, or making a contrasting argument using analogue and digital senses | Correct example of both bit and qubit |
| 1d | **Probabilities:** Incorrect<br>• Stated to be all ¼<br>• Other responses not of the form, e.g., $a_{00}^2$ or $|a_{00}|^2$ | **Probabilities:** e.g., $a_{00}^2$ | **Probabilities:** e.g., $|a_{00}|^2$ |
|  | **Probabilities add up to 1:** Not stated | **Probabilities add up to 1:** Addressed but not correct (None observed) | **Probabilities add up to 1:** Correct |
| 2a | Response is irrelevant (None observed) | Gives probabilities, but not correct (None observed) | Gives both correct probabilities |
| 2b | States such a gate cannot be made | Gives gate that yields desired state, but is not unitary | • Gives unitary gate that yields desired state<br>• States qualitatively that the gate "rotates" the state |
| 3a | Incorrect |  | Correct |
| 3b | Does not consider correct basis | Provides correct probabilities OR correct outcomes | Provides both correct probabilities and outcomes |
| 3c | Not enough information (calculates $\hat{H}|q\rangle$ with $|q\rangle = 0|0\rangle + 0|1\rangle$) | Provides correct state $\hat{H}|q\rangle$ only but not correct probability of measurement outcome | Provides correct state $\hat{H}|q\rangle$ and correct probability of measurement outcome |
| 3d | States such a gate cannot be made | Gives gate that yields desired state, but is not unitary | • Gives unitary gate that yields desired state<br>• States qualitatively that the gate "rotates" the state |
| 4a | True written (false is correct) | Correct answer with incorrect/incomplete explanation (e.g., disagrees with initialization to $|0\rangle$ instead of $2^N$ vs. $N$ states) | Correct answer and explanation (mentions classical computers have $2^N$, not $N$, possible states) |
| 4b | False written (true is correct) | Correct answer with incorrect/incomplete explanation | Correct answer and explanation (measurement collapses superposition to bit string) |
| 4c | True written (false is correct) | Correct answer with incorrect/incomplete explanation (e.g., circular answer) | Correct answer and explanation (qubits "keep track" of their own coefficients without need for classical registers) |
| 4d | True written (false is correct) | Correct answer with incorrect/incomplete explanation (e.g., does not mention that *not every* superposition state is entangled) | Correct answer and explanation (only certain superposition states, i.e., product states, are entangled) |
| 5 | Agrees with student 1 | Agrees with student 2 without explanation | Agrees with student 2 and gives explanation |
| 6 | Incorrect | | Correct |
| 7a | Gives incorrect probabilities (e.g., 50/50) | Percentages are correct but switched OR unspecified | Gives percentages for both right and wrong results |
| 7b | Does not adequately address either part of the question | Does not give all four $a_{ij}$ (e.g., "$a_{ij} = 1$" without specifying $i$ or $j$) OR does not mention quantum gates when considering how to obtain desired state (e.g., uses initialization or sampling ensemble of identical states, or says it's impossible) | Correctly gives all four $a_{ij}$ and mentions quantum gates |
| 7c | Gives incorrect probabilities | Percentages are correct but switched OR not specified | Gives percentages for correct and incorrect results |
| 8a | False written (true is correct) | Correct answer with incorrect/incomplete explanation | Correct answer and explanation (superposition is possible) |
| 8b | True written (false is correct) | Correct answer with incorrect/incomplete explanation | Correct answer and explanation (entanglement requires multiple qubits) |
| 8c | False written (true is correct) | Correct answer with incorrect/incomplete explanation | Correct answer and explanation (entanglement is possible) |
| 8d | True written (false is correct) without explanation | Incorrect/incomplete explanation (e.g., using CNOT to entangle states without recognizing that qubits interact in this process) | Correct answer and explanation (entanglement requires interaction between qubits) |
| 9a, 9b | Response is irrelevant (Only blank responses observed) | • Provides correct probabilities or correct outcomes<br>• Does not consider collapse | Provides both correct probabilities and outcomes |
| 9c | Gives incorrect probabilities (e.g., 50/50) | Does not consider collapse | Provides correct probabilities |
| 10a | Does not write bra states | Bra does not use complex conjugate | Correct |
| 10b | No matrix (e.g., gives ket) | Gives $a, b, c, d$ but not correct (e.g., $a = b = c = d$; assumes $\hat{I} = |q\rangle\langle q|$) | Correct $a, b, c, d$ |
| 10c | No matrix | Gives $a, b, c, d$ but not correct | Correct $a, b, c, d$ |
| 11 | Does not use outer products | Uses outer products, but does not write a correct expression | Correct, including $|0\rangle\langle+| + |1\rangle\langle-|$ or similar |

# Appendix C

Following are tables that break down the data in other potentially interesting ways. Tables 6-7 display results from both Physics classes separately, and we were able to collect data on long-term learning gains for the second Physics class, which are also shown in Table 7. Tables 8-9 presents results from the physics majors and non-physics majors in the QCQI course separately.
[†]In cases of $Pre \leq Post$, we follow the methodology for normalized change described by Marx and Cummings [56].

**Table 6.** Physics class 1, $N = 13$.

| Question | Pre | Post | Norm. Gain[†] | Effect size |
|---|---|---|---|---|
| 1a | 100% | 100% | 0 | 0 |
| 1b | 96% | 100% | 1.00 | 0.39 |
| 1c | 62% | 88% | 0.70 | 0.81 |
| 1d | 71% | 88% | 0.60 | 0.88 |
| 2a | 100% | 96% | -0.04 | -0.39 |
| 2b | 38% | 50% | 0.19 | 0.35 |
| 3a | 92% | 92% | 0.00 | 0.00 |
| 3b | 85% | 100% | 1.00 | 0.58 |
| 3c | 88% | 96% | 0.67 | 0.33 |
| 3d | 50% | 54% | 0.08 | 0.11 |
| 4a | 12% | 69% | 0.65 | 1.68 |
| 4b | 38% | 92% | 0.88 | 1.67 |
| 4c | 46% | 85% | 0.71 | 1.21 |
| 4d | 23% | 65% | 0.55 | 1.04 |
| 5 | 23% | 81% | 0.75 | 1.40 |
| 6 | 8% | 62% | 0.58 | 1.32 |
| 7a | 77% | 92% | 0.67 | 0.42 |
| 7b | 38% | 73% | 0.56 | 0.86 |
| 7c | 85% | 100% | 1.00 | 0.58 |
| 8a | 62% | 92% | 0.80 | 1.07 |
| 8b | 62% | 96% | 0.90 | 1.01 |
| 8c | 85% | 96% | 0.75 | 0.59 |
| 8d | 54% | 85% | 0.67 | 0.76 |

**Table 7.** Physics class 2, $N = 22$. We were able to collect data on long-term learning gains for some of these questions, which were asked on the final exam several months after the post-test. Gain and effect size are measured from the pre-test.

| Question | Pre | Post | Norm. Gain | Effect size | Final | Norm. Gain | Effect size |
|---|---|---|---|---|---|---|---|
| 1a | 98% | 100% | 1.00 | 0.31 | | | |
| 1b | 100% | 100% | 0 | 0 | | | |
| 1c | 64% | 93% | 0.81 | 0.95 | | | |
| 1d | 94% | 94% | 0.00 | 0.00 | | | |
| 2a | 100% | 100% | 0 | 0 | | | |
| 2b | 30% | 50% | 0.29 | 0.74 | | | |
| 3a | 95% | 100% | 1.00 | 0.31 | | | |
| 3b | 86% | 98% | 0.83 | 0.49 | | | |
| 3c | 64% | 89% | 0.69 | 0.67 | | | |
| 3d | 43% | 77% | 0.60 | 1.06 | | | |
| 4a | 41% | 89% | 0.81 | 1.22 | 98% | 0.96 | 1.67 |
| 4b | 64% | 100% | 1.00 | 1.19 | 98% | 0.93 | 1.08 |
| 4c | 57% | 91% | 0.79 | 1.14 | 100% | 1.00 | 1.76 |
| 4d | 75% | 95% | 0.82 | 0.62 | 95% | 0.80 | 0.60 |
| 5 | 55% | 91% | 0.80 | 0.93 | 73% | 0.57 | 0.79 |
| 6 | 36% | 91% | 0.86 | 1.38 | | | |
| 7a | 82% | 100% | 1.00 | 0.80 | | | |
| 7b | 61% | 80% | 0.47 | 0.48 | | | |
| 7c | 89% | 100% | 1.00 | 0.54 | | | |
| 8a | 75% | 91% | 0.64 | 0.44 | | | |
| 8b | 77% | 95% | 0.80 | 0.58 | | | |
| 8c | 82% | 98% | 0.88 | 0.61 | | | |
| 8d | 66% | 89% | 0.67 | 0.61 | | | |
| 9a | 82% | 100% | 1.00 | 0.72 | | | |
| 9b | 82% | 100% | 1.00 | 0.72 | | | |
| 9c | 91% | 95% | 0.50 | 0.20 | | | |
| 10a | 91% | 95% | 0.50 | 0.27 | | | |
| 10b | 70% | 77% | 0.23 | 0.21 | 85% | 0.49 | 0.43 |
| 10c | 66% | 89% | 0.67 | 0.70 | | | |
| 10d | 57% | 89% | 0.74 | 0.82 | | | |
| 11 | 30% | 73% | 0.61 | 1.03 | | | |

**Table 8.** The data presented in the QCQI course in Table 2, with only the physics majors shown ($N = 11$) and the non-physics majors ($N = 7$). Pre-test and post-test average scores, normalized gain, and effect size as measured by Cohen's $d$ are presented. Student data for the pre- and post-test in each case are matched. An additional column is included containing (unmatched) data from all students who completed the post-test ($N = 12$). †In cases of $Pre \leq Post$, we follow the methodology for normalized change described by Marx and Cummings.

| Question | Pre (N=11) | Post (matched) (N=11) | Norm. Gain† | Effect size | Post (unmatched) (N=12) |
|---|---|---|---|---|---|
| 1a | 100% | 91% | -0.09 | -0.64 | 92% |
| 1b | 100% | 95% | -0.05 | -0.43 | 96% |
| 1c | 59% | 82% | 0.56 | 0.64 | 79% |
| 1d | 82% | 93% | 0.63 | 0.63 | 94% |
| 2a | 100% | 100% | 0 | 0 | 100% |
| 2b | 59% | 77% | 0.44 | 0.56 | 75% |
| 3a | 100% | 91% | -0.09 | -0.43 | 83% |
| 3b | 95% | 100% | 1.00 | 0.43 | 100% |
| 3c | 77% | 91% | 0.60 | 0.48 | 92% |
| 3d | 50% | 68% | 0.36 | 0.56 | 63% |
| 4a | 50% | 86% | 0.73 | 0.86 | 88% |
| 4b | 73% | 86% | 0.50 | 0.37 | 88% |
| 4c | 50% | 95% | 0.91 | 1.36 | 96% |
| 4d | 64% | 86% | 0.63 | 0.54 | 88% |
| 5 | 45% | 77% | 0.58 | 0.68 | 79% |
| 6 | 45% | 82% | 0.67 | 0.78 | 83% |
| 7a | 73% | 100% | 1.00 | 0.83 | 100% |
| 7b | 50% | 82% | 0.64 | 0.80 | 79% |
| 7c | 77% | 100% | 1.00 | 0.78 | 100% |
| 8a | 91% | 86% | -0.05 | -0.17 | 88% |
| 8b | 95% | 100% | 1.00 | 0.43 | 100% |
| 8c | 86% | 93% | 0.50 | 0.34 | 94% |
| 8d | 70% | 91% | 0.70 | 0.57 | 92% |
| 9a | 82% | 100% | 1.00 | 0.64 | 100% |
| 9b | 82% | 100% | 1.00 | 0.64 | 100% |
| 9c | 95% | 100% | 1.00 | 0.43 | 100% |
| 10a | 82% | 100% | 1.00 | 0.76 | 100% |
| 10b | 64% | 68% | 0.13 | 0.12 | 71% |
| 10c | 59% | 80% | 0.50 | 0.58 | 79% |
| 11 | 27% | 68% | 0.56 | 0.94 | 63% |

**Table 9.** The data presented in the QCQI course in Table 2, with only the non-physics majors shown ($N = 7$). Pre-test and post-test average scores, normalized gain, and effect size as measured by Cohen's $d$ are presented. Student data for the pre- and post-test in each case are matched. An additional column is included containing (unmatched) data from all students who completed the post-test ($N = 16$). †In cases of $Pre \leq Post$, we follow the methodology for normalized change described by Marx and Cummings.

| Question | Pre (N=7) | Post (matched) (N=7) | Norm. Gain† | Effect size | Post (unmatched) (N=16) |
|---|---|---|---|---|---|
| 1a | 100% | 100% | 0 | 0 | 100% |
| 1b | 100% | 89% | -0.11 | -0.54 | 95% |
| 1c | 57% | 100% | 1.00 | 1.35 | 84% |
| 1d | 93% | 93% | 0 | 0 | 92% |
| 2a | 100% | 100% | 0 | 0 | 100% |
| 2b | 43% | 50% | 0.13 | 0.19 | 41% |
| 3a | 100% | 100% | 0 | 0 | 100% |
| 3b | 71% | 100% | 1.00 | 0.83 | 94% |
| 3c | 79% | 100% | 1.00 | 0.77 | 94% |
| 3d | 43% | 71% | 0.50 | 0.93 | 66% |
| 4a | 50% | 86% | 0.71 | 1.06 | 91% |
| 4b | 43% | 93% | 0.88 | 1.45 | 91% |
| 4c | 64% | 86% | 0.60 | 0.67 | 80% |
| 4d | 71% | 100% | 1.00 | 1.51 | 84% |
| 5 | 57% | 86% | 0.67 | 0.62 | 91% |
| 6 | 29% | 86% | 0.80 | 1.31 | 81% |
| 7a | 86% | 86% | 0 | 0 | 91% |
| 7b | 43% | 71% | 0.50 | 0.68 | 72% |
| 7c | 57% | 100% | 1.00 | 1.13 | 100% |
| 8a | 79% | 100% | 1.00 | 0.77 | 100% |
| 8b | 79% | 100% | 1.00 | 0.77 | 97% |
| 8c | 71% | 86% | 0.50 | 0.44 | 88% |
| 8d | 25% | 86% | 0.81 | 1.52 | 75% |
| 9a | 100% | 100% | 0 | 0 | 94% |
| 9b | 100% | 100% | 0 | 0 | 100% |
| 9c | 86% | 86% | 0 | 0 | 94% |
| 10a | 57% | 100% | 1.00 | 1.76 | 97% |
| 10b | 57% | 64% | 0.17 | 0.15 | 72% |
| 10c | 29% | 64% | 0.50 | 0.93 | 58% |
| 11 | 21% | 57% | 0.45 | 0.76 | 34% |